\documentclass[a4paper,11pt]{article}
\pdfoutput=1
\usepackage{jcappub}

\usepackage{amsthm,graphicx}
\usepackage{epsfig}
\usepackage{latexsym, amssymb} 
\usepackage{amsmath}
\usepackage[normalem]{ulem}
\usepackage{xcolor}

\usepackage{rotating}

\usepackage{natbib}

\let\mr=\mathrm

\newcommand{\bq}{\begin{equation}}
\newcommand{\eq}{\end{equation}}
\newcommand{\LCDM}{\Lambda\mathrm{CDM}}
\newcommand{\hpM}{h\,\mathrm{Mpc}^{-1}}
\newcommand{\Mpc}{\mathrm{Mpc}^{-1}}
\def\equationautorefname~#1\null{equation~(#1)\null}
\def\sectionautorefname~#1\null{Section~#1\null}
\def\figureautorefname~#1\null{Figure~#1\null}

\begin{document}
\title{Probing inflation with large-scale structure data: the contribution of information at small scales}

\author[1]{
Ivan Debono}
\affiliation{Paris Centre for Cosmological Physics, Universit\'e de Paris,
CNRS, Astroparticule et Cosmologie\\ F-75006 Paris, France}
\emailAdd{mail@ivandebono.eu}

\abstract{Upcoming full-sky large-scale structure surveys such as Euclid can probe the primordial Universe. Using the specifications for the Euclid survey, we estimate the constraints on the inflation potential beyond slow-roll. We use mock Euclid and Planck data from fiducial cosmological models using the Wiggly Whipped Inflation (WWI) framework, which generates features in the primordial power spectrum. We include Euclid cosmic shear and galaxy clustering, with two setups (Conservative and Realistic) for the non-linear cut-off. We find that the addition of Euclid data gives an improvement in constraints in the WWI potential, with the Realistic setup providing marginal improvement over the Conservative for most models. This shows that Euclid may allow us to identify oscillations in the primordial spectrum present at intermediate to small scales.}

% Keywords
%\keyword{cosmology; inflation; weak lensing; cosmic microwave background}

\maketitle

%%%%%%%%%%%%%%%%%%%%%%%%%%%%%%%%%%%%%%%%%%

%%%%%%%%%%%%%%%%%%%%%%%%%%%%%%%%%%%%%%%%%%

\section{Introduction}

The last two decades have seen huge advances in the measurement of cosmological parameters. Full-sky surveys can probe physics at the largest cosmological scales, and the next generation of probes may provide answers to the open questions on the Concordance Model. This model---commonly referred to as  $\Lambda$ Cold Dark Matter ($\LCDM$)---can fit different astrophysical datasets with just six parameters describing the mass--energy content of the Universe and the initial conditions. The content consists of baryons, CDM and a cosmological constant or constant dark energy. The initial conditions are parametrized by a phenomenological fit with a smooth primordial power spectrum. 

Despite the success of the Concordance Model, there are three big open questions in modern cosmology.
\begin{enumerate}

\item The nature of dark matter, which constitutes the bulk of the matter content. 

\item The component causing the accelerated expansion of the Universe. This may be a cosmological constant ($\Lambda$, or some additional component known as dark energy, which may be dynamical, with a redshift-dependent equation of state (parametrized by some expression for $w$, e.g., $w=w_0+w_a(1-a)$ or it may be a constant.

\item Conditions in the very early Universe. The Theory of Inflation is well-established, and has been confirmed with remarkable precision by a succession of cosmic microwave background (CMB) probes. WMAP~\cite{WMAP:2003,WMAP9}
provided conclusive evidence for inflation. Planck~\cite{Planck2018:params,Planck2018:inflation} conclusively excluded a scale-invariant primordial power spectrum. What is the form of this power spectrum beyond its main shape and amplitude? Does it contain features? If so, at which scales do they occur? What is the inflaton potential producing this power spectrum?
\end{enumerate}
 
The data are compatible with a Universe filled with dark matter and cosmological constant, with a smooth primordial power spectrum. However, do not exclude dynamical dark energy. Nor do they exclude features in the primordial power spectrum.

We focus our attention on measuring possible features in the primordial power spectrum. This paper is a companion to ref.~\cite{Debono2020}, in which the authors (including the present author) quantified the projected constraints from Euclid in the presence of features in the primordial power spectrum, and the improvement provided by Euclid over Planck in measuring inflation parameters. Here we show how the inclusion of information from large-scale structure at even smaller scales can improve constraints from Euclid.

We use Wiggly Whipped Inflation (WWI; \cite{Hazra2014}), which can generate a variety of primordial power spectra with features at different cosmological scales. Since Euclid data are not yet available, we have to simulate them. We use the Planck best-fitting Wiggly Whipped Inflation models to create fiducial cosmologies and thus data for Planck and Euclid. Then we use Markov chain Monte Carlo (MCMC) simulations for cosmological parameter estimation.
\section{Primordial Physics}
\label{sec:PrimordialPhysics}

The large-scale structure we observe today in the Universe was seeded by primordial quantum perturbations which originated and evolved during the inflationary epoch. The shape of the primordial power spectrum describing these perturbations depends on the inflation potential.

The simplest primordial scalar power spectrum is a power law with the following phenomenological form:
 \bq 
 P_\mathrm{S}(k)=A_\mathrm{s}\left(\frac{k}{k_{0}}\right)^{n_{\mathrm{s}}-1} \ ,
 \eq 
 where $A_\mathrm{s}$ is the amplitude and $n_{\mathrm{s}}$ is the tilt
 of the spectrum of primordial perturbations~\cite{Kosowsky:1995, Bridle:2003}. This is used in the Concordance Model of cosmology.  The scale-invariant power spectrum with $n_{\mathrm{s}}=1$ is now firmly excluded by observation \cite{Planck-Collaboration:2013aa,Planck-Collaboration2015aa,Planck2018:params,Planck2018:inflation}. 

This power spectrum is featureless. Features can be described by variations of the power-law parametrization. Broad features can be parametrized by logarithmic derivatives of the tilt (running and running-of-running), or by local and non-local wiggles in the power spectrum. To date, the only properties which have been established with any statistical significance are the amplitude and the tilt of the primordial power spectrum. 

In different reconstructions, primordial features at particular scales have been found to address
tensions between data sets present with $\LCDM$
~\cite{Hannestad_2001,Tegmark_2002,Bridle:2003,Mukherjee_2003,Shafieloo_2004,Kogo_2005,Leach:2005av,TocchiniValentini:2005ja,Shafieloo_2008,Nicholson_2009,Paykari_2010,Gauthier_2012,Hlozek_2012,Hazra_2013,Dorn_2014,Hazra_2014,Hazra2014bJCAP,Hunt_2014}.
 
3D surveys such as Euclid can provide joint estimates with CMB data. In ref.~\cite{Debono2020} and this companion paper, we use the MCMC method to forecast the constraints on possible oscillations in the primordial spectrum. Instead of a parametric modification to the power-law spectrum, we model the existence of such features directly from inflation theory.
%%%

\subsection{The Inflationary Potential}
\label{sec:InflationaryPotential}

In this section we give the essential details of the inflationary potentials used in our cosmological models. Further details are found in ref.~\cite{Debono2020}, and references therein.
Wiggly Whipped Inflation was first proposed in~\cite{Hazra2014}. It is an extension of the Whipped Inflation model introduced in ref.~\cite{Hazra2014b}. Its most distinctive feature is the presence of wiggles in the primordial power spectrum (hence the name). Both Wiggly Whipped and Whipped Inflation belong to the class of models with a large field inflaton potential. 

We consider two WWI potentials, which we call Wiggly Whipped Inflation (hereafter, WWI potential) and Wiggly Whipped Inflation Prime (WWIP potential).

The WWI potential is defined by:
\begin{equation}
V({\phi})=V_{i} \left(1-\left(\frac{\phi}{\mu}\right)^{p}\right)+\Theta(\phi_{\mr T}-\phi) V_{i}\left(\gamma (\phi_{\mr T}-\phi)^{q}+\phi_{0}^q\right),~\label{eq:equation-WWI}
\end{equation}
where $V_{S}(\phi)=V_{i} \left(1-\left(\frac{\phi}{\mu}\right)^{p}\right)$ has two parameters, $V_{i}$ and $\mu$. The parameter $\mu$ and the index $p$ determine the spectral tilt $n_{\mr s}$ and the tensor-to-scalar ratio $r$.
We set $p=4$ and $\mu=15~M_\mathrm{P}$, where $M_\mathrm{P}=1$ is the reduced Planck mass, such that $n_{\mr s}\sim0.96$ and $r\sim{\cal O}(10^{-2})$.
The transition and discontinuity occur at the field value $\phi_{\mr T}$.
If $\gamma=0$ and $\phi_{0}=0$, a featureless primordial power spectrum is obtained. The Heaviside Theta function $\Theta(\phi_{\mr T}-\phi)$ is modelled numerically by a Tanh step
($\frac{1}{2}\left[1+{\tanh}[{(\phi-\phi_{\mr T})}/{\delta}]\right]$) and thus introduces a new extra parameter $\delta$.

The WWIP potential is described in ref.~\cite{Hazra2016}. It is defined by:
\begin{align}
V({\phi})=\, & \Theta(\phi_{\mr T}-\phi) V_{i}
\left(1-\exp\left[-\alpha\kappa\phi\right]\right)\nonumber \\
&+\Theta(\phi-\phi_{\mr T})
V_{ii}\left(1-\exp\left[-\alpha\kappa(\phi-\phi_{0})\right]\right)\, . ~\label{eq:equation-WWIP}
\end{align}

We set $\alpha=\sqrt{2/3}$. In our convention, $\kappa^2=8\pi G$ is equal to $1$, where $G$ is the gravitational constant.

%%% 

%
\section{Method}
\label{sec:Method}

Our forecasts use the MCMC technique, with mock data from fiducial cosmological models. The Euclid likelihoods used in this paper are described in detail in ref.~\cite{Debono2020}.

We compute mock data from a fiducial cosmology following the method defined in ref.~\cite{Sprenger2018}. We carry out three MCMC forecasts for each cosmological model, for a total of $24$~forecasts: 
\begin{enumerate}
\item	Simulated Planck CMB data alone (shown in red in the triangle plots); 
\item Joint Euclid Conservative galaxy clustering $+$ Euclid Conservative cosmic shear $+$ simulated Planck CMB data (shown in blue);
\item	Joint Euclid Realistic galaxy clustering $+$ Euclid Realistic cosmic shear $+$ simulated Planck CMB data (shown in green).
\end{enumerate}

The details for Planck and Euclid Conservative (galaxy clustering and cosmic shear) are described in ref.~\cite{Debono2020}.
\subsection{The Non-Linear Theoretical Uncertainty: `Conservative' and `Realistic' Setups}

The difference between Euclid Conservative and Realistic is in the cutoff at non-linear scales, following ref.~\cite{Sprenger2018}. This defines a cutoff $k_\mathrm{NL}$. All theoretical uncertainties up to this wavenumber are ignored, while all the information above it is discarded. The redshift dependence of non-linear effects is parametrized BY:
\bq
\label{eq:nonlinear}
k_\mathrm{NL}(z)=k_\mathrm{NL}(0) (1+z)^{2/(2+n_\mathrm{s})} \ .
\eq

This gives us two frameworks for modelling the theoretical error. The first is a `realistic' case where the parametrization of the error is used up to large wavenumbers, and an increasing relative error function gradually suppressed the information from small scales. The second is a `conservative' case where the same error function is used, but with a sharp cut-off. We will henceforth capitalize these two terms for clarity: Realistic and Conservative.

In ref.~\cite{Debono2020}, the parameters for galaxy clustering and cosmic shear forecast correspond to the Conservative setup. In this paper, we show both Conservative and Realistic. 

The differences are the following:
\begin{itemize} 
\item Conservative galaxy clustering: We use a cut-off on large wavelengths at $k_\mr{min}=0.02~\Mpc$. This eliminates scales which are bigger than the bin width or which violate the small-angle approximation. On small wavelengths, we use a theoretical uncertainty with $k_\mr{NL}(0)=0.2 \hpM$.
\item Realistic galaxy clustering: The same formulation, but with $k_\mathrm{max} = 10 \, \hpM$
\end{itemize}

\begin{itemize}
\item Conservative cosmic shear: We include multipoles from $\ell_\mathrm{min}=5$ up to a bin-dependent non-linear cut-off given by $k_\mr{NL}(0)=0.5 \hpM$
\item Realistic cosmic shear: the same, but with $k_\mathrm{NL}(0)= 2\, \hpM$.
\end{itemize}

%%%

%

\subsection{Fiducial Cosmology and WWI Models}
\label{sec:Cosmology}

We assume a Friedmann--Robertson--Walker cosmology with a flat spatial geometry. 
The background $\LCDM$ cosmology is parametrized by: the baryon density $\omega_{\mathrm{b}}=\Omega_\mathrm{b} h^2$, the cold dark matter density $\omega_{\mathrm{cdm}}=\Omega_\mathrm{cdm} h^2$, the Hubble parameter via the peak scale parameter $100\theta_{\mathrm{s}}$, and the optical depth to reionization $\tau_{\mathrm{reio}}$. 
We use the following values for all our models:
 $\omega_{\mathrm{b}}= 2.21\times 10^{-2}$,
 $\omega_{\mathrm{cdm}}=0.12$,
$100\theta_{\mathrm{s}}=1.0411$, and 
$\tau_{\mathrm{reio}}=0.09$.
Our models include massive neutrinos. We assume three neutrino species, with the total neutrino mass split according to a normal hierarchy. All neutrino parameters are kept fixed. The sum of the neutrino masses $M_\mr{total}=0.06$~eV, and the number of effective neutrino species in the early Universe $N_\mr{eff}=3.046$. 

Besides the four parameters for the $\LCDM$ background, we have the inflationary potential parameters. Table~\ref{tab:InflationModels} shows their fiducial values.
We use five free parameters for WWI, and three for WWIP. The parameter spaces of the cosmological models used in our MCMC simulations therefore include seven parameters for
WWI:$\{\omega_{\mathrm{b}},\, \omega_{\mathrm{cdm}},\, 100\theta_{\mathrm{s}},\, \tau_{\mathrm{reio}},\,$\linebreak$
\ln(10^{10}V_0),\, \phi_{0},\, \gamma,\, \phi_\mathrm{T},\, \ln\delta \}$, and nine parameters for
WWIP: $\{ \omega_{\mathrm{b}},\, \omega_{\mathrm{cdm}},\, 100\theta_{\mathrm{s}}, \,\tau_{\mathrm{reio}},\,$\linebreak$ \ln(10^{10}V_0),\, \phi_{0},\, \phi_\mathrm{T}
 \}$.

For the WWI potential, we consider five models: one featureless power spectrum (called WWI: Featureless), and four with different types of features at different scales corresponding to local and global best fits to the Planck data. We call these WWI-[A, B, C, D], following the naming convention in ref.~\cite{Hazra2016}. 

For the WWIP potential, we consider three models. Two have features: the Planck global best-fitting spectrum (WWIP: Planck-best-fit) \cite{Hazra2016}, and a spectrum within the 95 per cent Planck confidence limits (WWIP: Small-scale-feature) with wiggles extending to smaller scales. As for WWI, we include one spectrum without features (WWIP: Featureless).

The two featureless spectra are obtained by fixing $\phi_{0}=0$, $\gamma=0$ for WWI, and $\phi_{0}=0$ for WWIP. 
%%% 

\begin{table}[tbp]
\centering

\begin{tabular}{l c c c c c}
\hline
\textbf{Model} & \textbf{$\ln(10^{10}V_0)$} & \textbf{$\phi_{0}$ }& \textbf{$\gamma$} & \textbf{$\phi_\mathrm{T}$} & \textbf{$\ln\delta$} \\
\hline
WWI: Featureless & $1.73 $ &$ 0 $& $0 $& -- & -- \\ 
WWI--A           & $1.73$  & $0.0137$ & $0.019$ & $7.89$ & $-4.5$ \\
WWI--B           & $1.75 $ & $0.0038$ & $0.04$ & $7.91$ & $-7.1$ \\
WWI--C           & $1.72 $ & $0.0058$ & $0.02$ & $7.91$ & $-6$ \\
WWI--D           & $1.76 $ & $0.003$ & $0.033$ & $7.91$ & $-11$ \\
WWIP: Featureless & $0.282$     & 0   & -- & -- & -- \\
WWIP: Planck-best-fit & $0.282$     & $0.11$ & -- & $4.51$ & -- \\
WWIP: Small-scale-feature & $0.3$ & $0.18$ &-- & $4.5$ & -- \\
\hline
\end{tabular}
\caption{
Parameter values for the inflationary potential parameters used to obtain the fiducial primordial power spectra.\label{tab:InflationModels}
}
\end{table}

We use the \textsc{bingo} package \cite{Bingo} to compute the primordial power spectrum from the inflation models. This is used within
the MCMC sampler \textsc{montepython} \cite{MontePython3} with the Boltzmann solver \textsc{class} \cite{CLASS}. We include both cosmic shear and galaxy clustering, using the Conservative and Realistic likelihoods described in ref.~\cite{Sprenger2018}.
The non-linear part of the spectrum is calculated using the \textsc{halofit} formula \cite{Takahashi2012,Bird2012}.
%%%

%
\section{Results}
\label{sec:Results}

We fit the theoretical sampled angular power spectra and matter power spectra to the fiducial data of the corresponding models, and thus obtain Planck-only and joint Euclid+Planck constraints. Euclid includes galaxy clustering and cosmic shear. %%%

\begin{figure}[!htb]
\centering
	\includegraphics[width=\columnwidth]{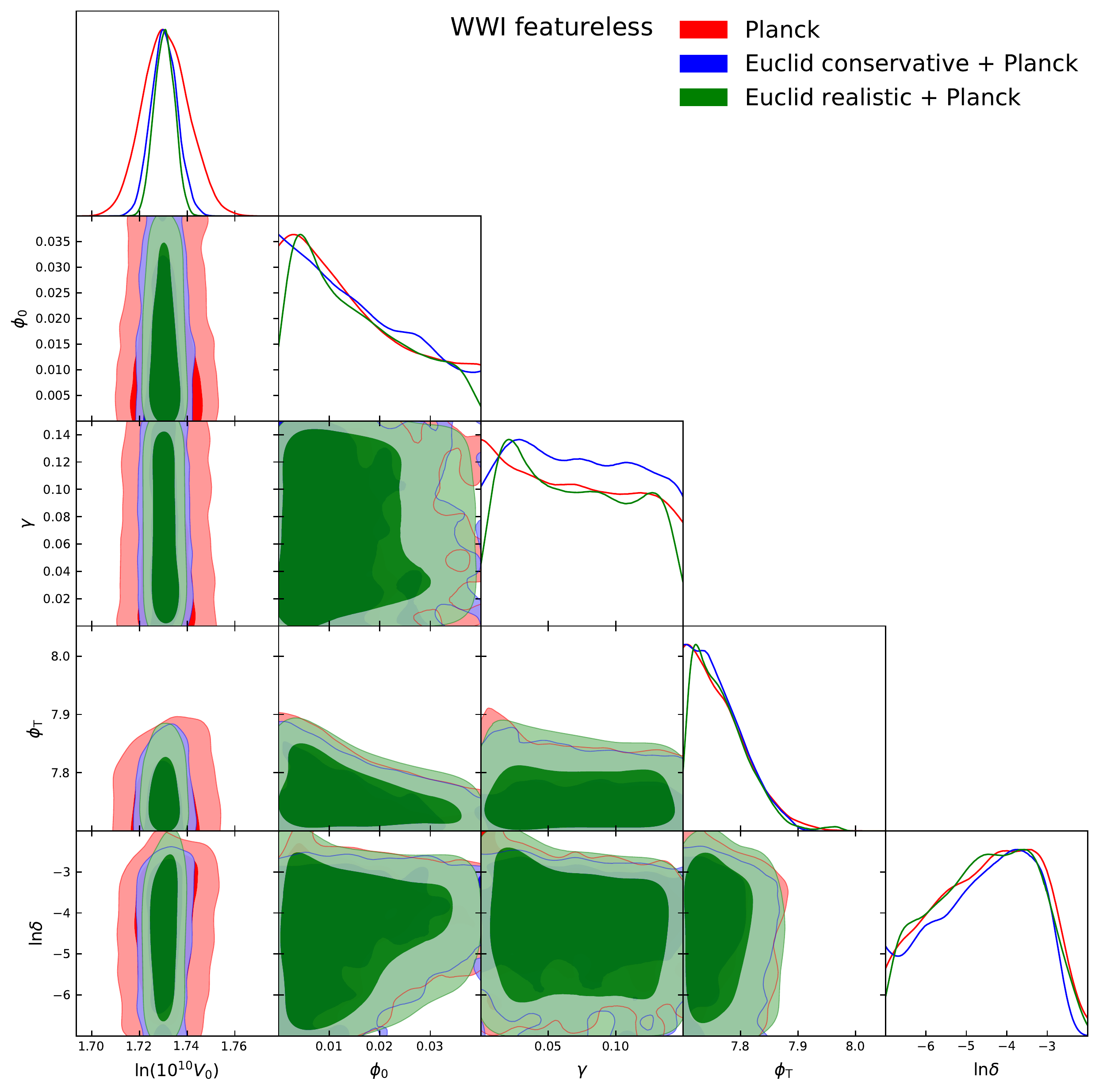}
    \caption{One-dimensional posteriors and marginalized contours ($1\sigma$ and $2\sigma$) for the inflation parameters in the WWI: Featureless model. 
    The addition of Euclid data results in a significant improvement in constraints for the amplitude parameter, and there is only slight improvement with Euclid Realistic compared to Euclid Conservative.
    \label{fig:WWIFeatureless}}
\end{figure}

\begin{figure}[!htb]
\centering
	\includegraphics[width=\columnwidth]{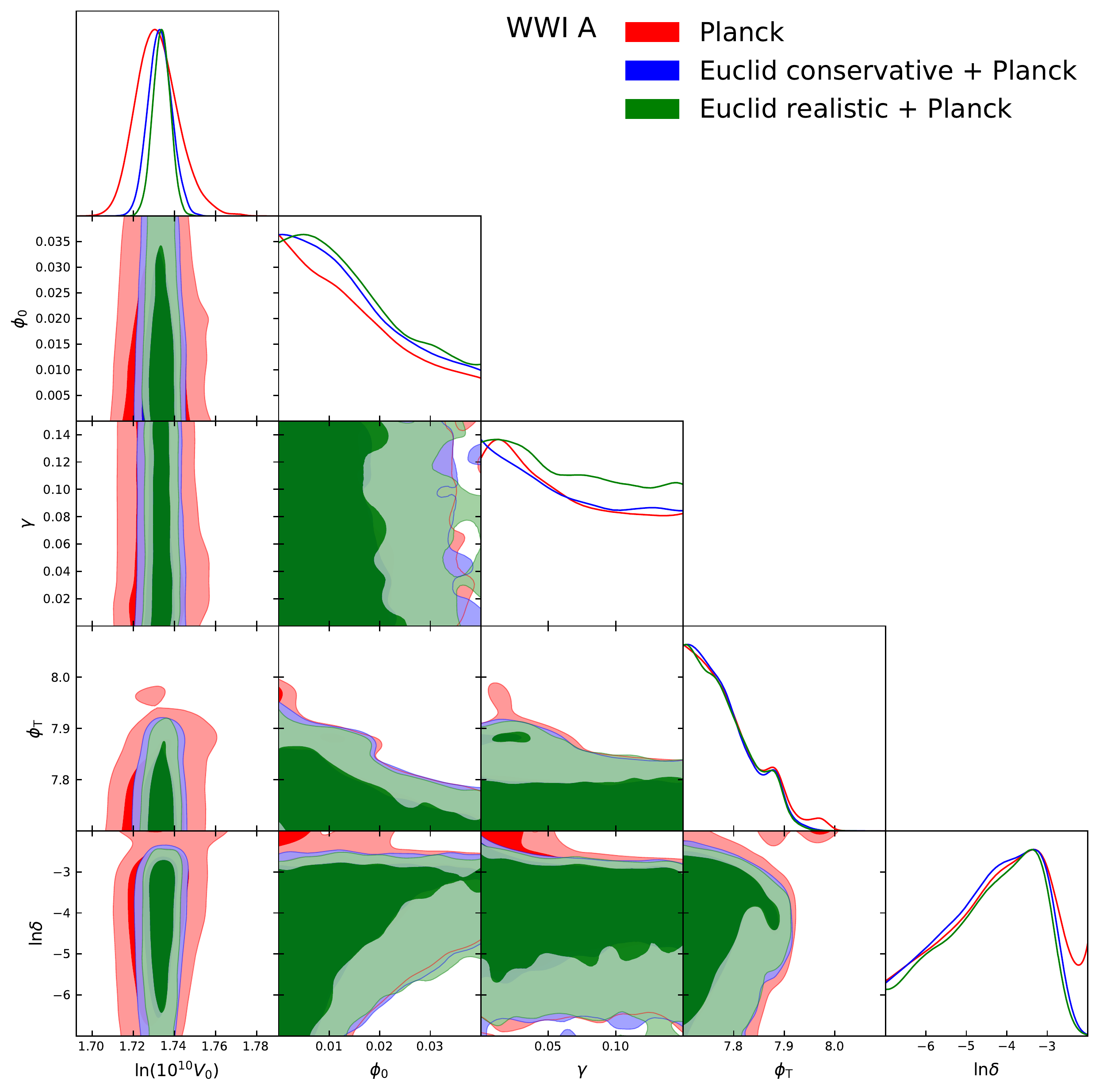}
    \caption{One-dimensional posteriors and marginalized contours for the inflation parameters in the WWI--A model.
     Improvement in the constraints is most evident in the amplitude parameter $V_0$. The constraints from Euclid Realistic are slightly better than Euclid Conservative. 
    \label{fig:WWIA}}
\end{figure}

\begin{figure}[!htb]
\centering
	\includegraphics[width=\columnwidth]{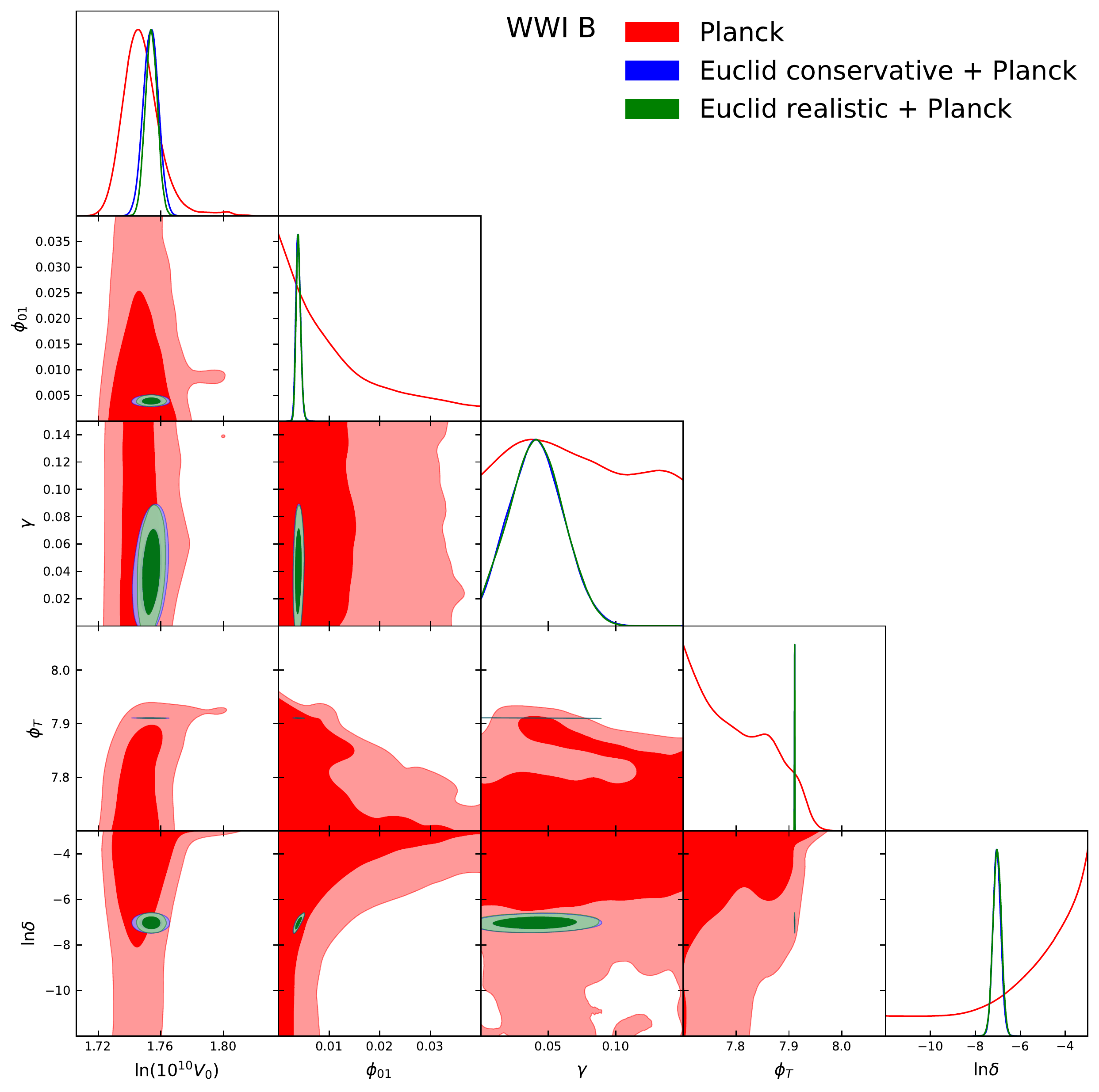}
    \caption{One-dimensional posteriors and marginalized contours for the inflation parameters in the WWI--B model. 
    There is significant improvement in constraints for all inflation parameters with the addition of Euclid data, but little difference between Euclid Conservative and Realistic.
    \label{fig:WWIB}}
\end{figure}

\begin{figure}[!htb]
\centering
	\includegraphics[width=\columnwidth]{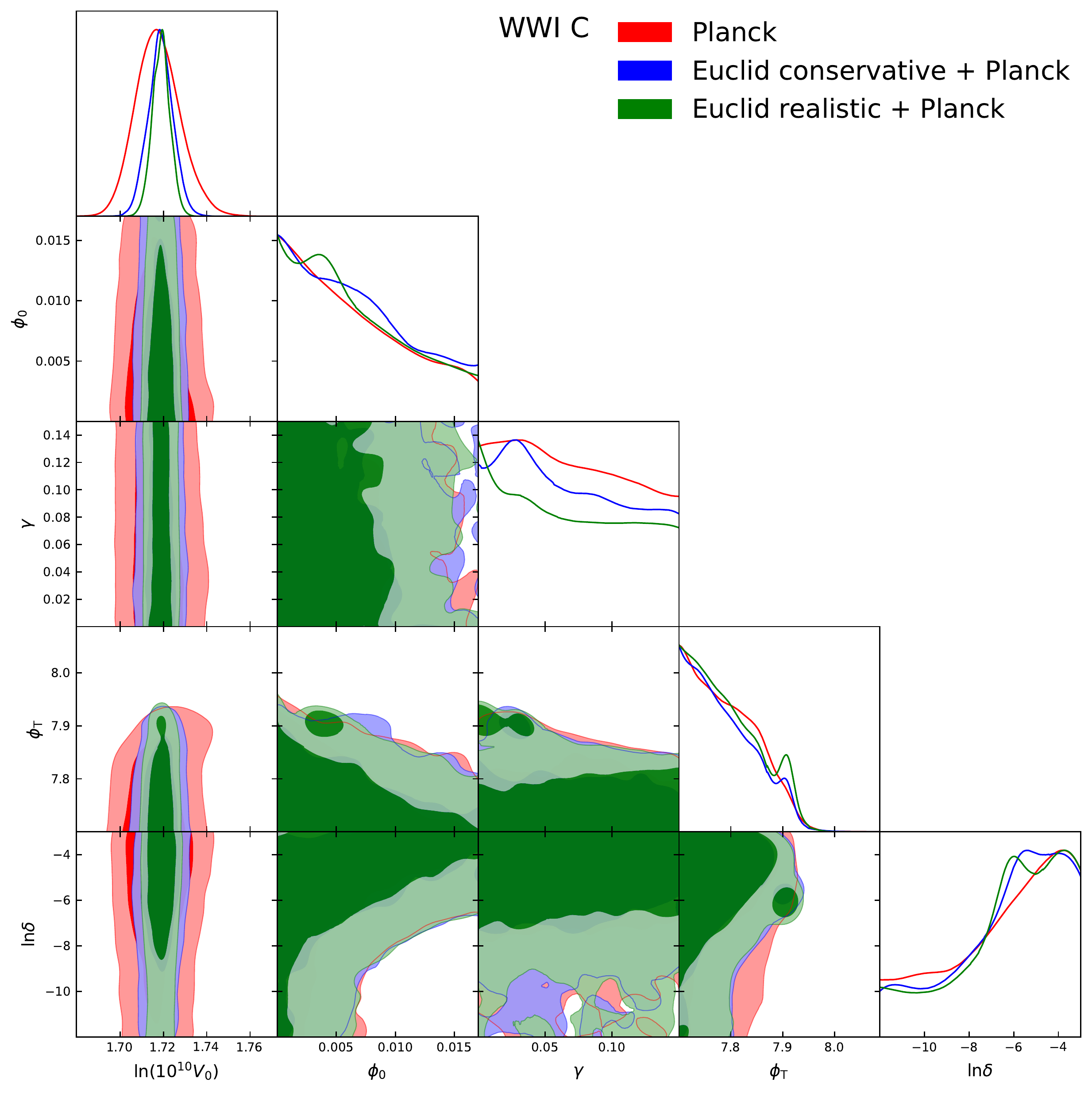}
    \caption{One-dimensional posteriors and marginalized contours for the inflation parameters in the WWI--C model. 
    As with WWW--A, there is some improvement when Euclid Realistic is used.
    \label{fig:WWIC}}
\end{figure}

\begin{figure}[!htb]
\centering
	\includegraphics[width=\columnwidth]{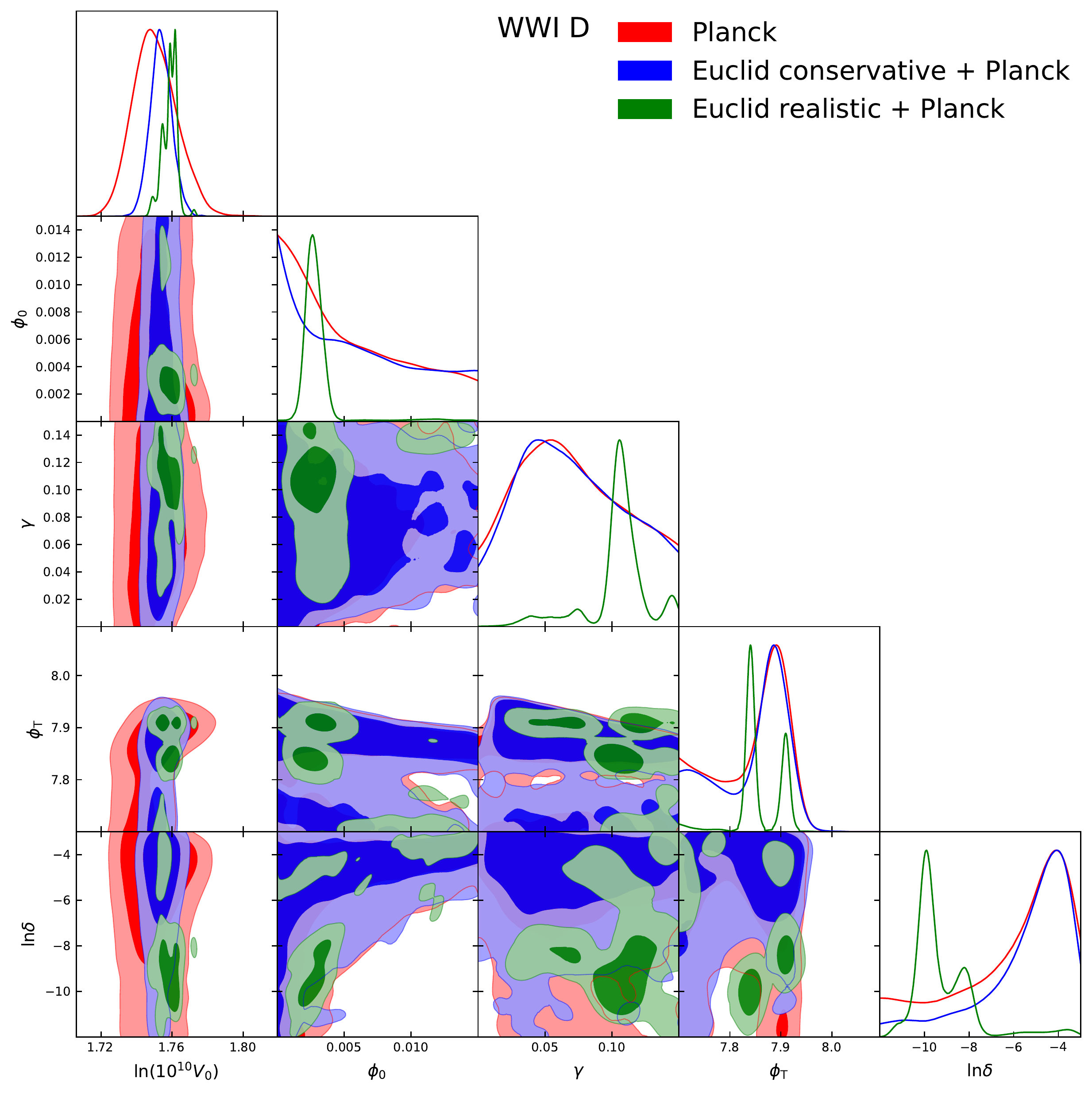}
    \caption{One-dimensional posteriors and marginalized contours for the inflation parameters in the WWI--D model. We note a significant improvement from Euclid Realistic over Euclid Conservative for all parameters.
    \label{fig:WWID}}
\end{figure}

The featureless WWI model gives a smooth primordial spectrum with a spectral tilt of $0.96$, corresponding to the current best estimate for the power-law spectrum. The $1\sigma$ and $2\sigma$ constraints on inflation potential parameters are shown in Figure~\ref{fig:WWIFeatureless}. Aside from $V_0$, we do not obtain any improvement by adding Euclid data. If the primordial power spectrum follows a power law, Euclid is not likely to rule out any large-scale power suppression (induced by $\gamma$) or oscillations (induced by $\phi_0$) with higher statistical significance than Planck has already done. Planck already rules out wiggles at small scales. If the real data are close to a featureless power spectrum, Euclid is not expected to rule out potentials already ruled out by Planck, regardless of whether the Euclid Conservative or Realistic setup is used. 
%%%

%
WWI-A to D produce suppression of the power spectrum at large scales, and wiggles at intermediate scales (which are probed by Euclid). For A to C, they die out at small scales, while they persist at smaller scales for WWI-D. As we shall see later, this distinction determines the relative contribution of information from smaller scales to the constraints.

We show inflation potential parameters for WWI-A in Figure~\ref{fig:WWIA}. 
We do not observe any improvement in the potential parameters with either of the Euclid setups, except for an improvement in $V_0$ (the amplitude of the primordial spectrum). Euclid cannot constrain the power spectrum at the largest scales better than Planck, since the Euclid measurement error at these scales is dominated by statistical uncertainties due to cosmic variance. Euclid can marginally tighten the bounds on the frequency of the oscillation by constraining $\ln\delta$.

The WWI-B, C and D fiducial models have wiggles in the primordial spectrum at intermediate to smaller scales ($\sim$0.1~$\mathrm{Mpc}^{-1}$), which fall within the high signal-to-noise region of both Planck and Euclid. 
The posteriors and marginalized contours for WWI-B, WWI-C and WWI-D are shown in Figure~\ref{fig:WWIB}--\ref{fig:WWID}. 

There is a remarkable improvement in constraints for WWI-B when Euclid is combined with Planck, but only minimal improvement when the Realistic setup is used. We expect Euclid data in combination with Planck to provide substantial evidence for WWI-B if it fits the data as well as $\Lambda$CDM. 
%%%

%
In WWI-C, there are wiggles at intermediate scales, which decay at smaller scales ($k\sim10^{-2}\mathrm{Mpc}^{-1}$). The limited overlap with cosmological scales probed by Euclid reduces the chances of a detection of these features. Since the wiggles die out before the nonlinear regime, using the Realistic setup only provides a small improvement.

WWI-D is where we observe most clearly the effect of including information at smaller scales. Euclid Conservative improves the constraints on the inflationary parameters compared to Planck-only results, and we obtain a significant improvement when we use the Realistic setup. The detection of features is unlikely with Euclid Conservative, but it becomes possible with the Realistic setup. At the scales where Planck and Euclid coverage overlap, the features WWI-D have the smallest amplitude out of the WWI models (see Figures 1 and 2 in ref.~\cite{Debono2020}). However, unlike the wiggles in the other WWI models, the oscillations in WWI-D persist at smaller scales. Since the Euclid Conservative wavenumber cutoff limits the information from smaller scales, it cannot resolve the high frequency oscillations since they are binned and averaged out in the observed power spectrum.  Euclid Realistic includes these scales, which explains why it can provide such a significant improvement on the Conservative setup. The challenge is therefore to model the non-linear power spectrum as accurately as possible.

\begin{figure}[!htb]
\centering
	\includegraphics[width=\columnwidth]{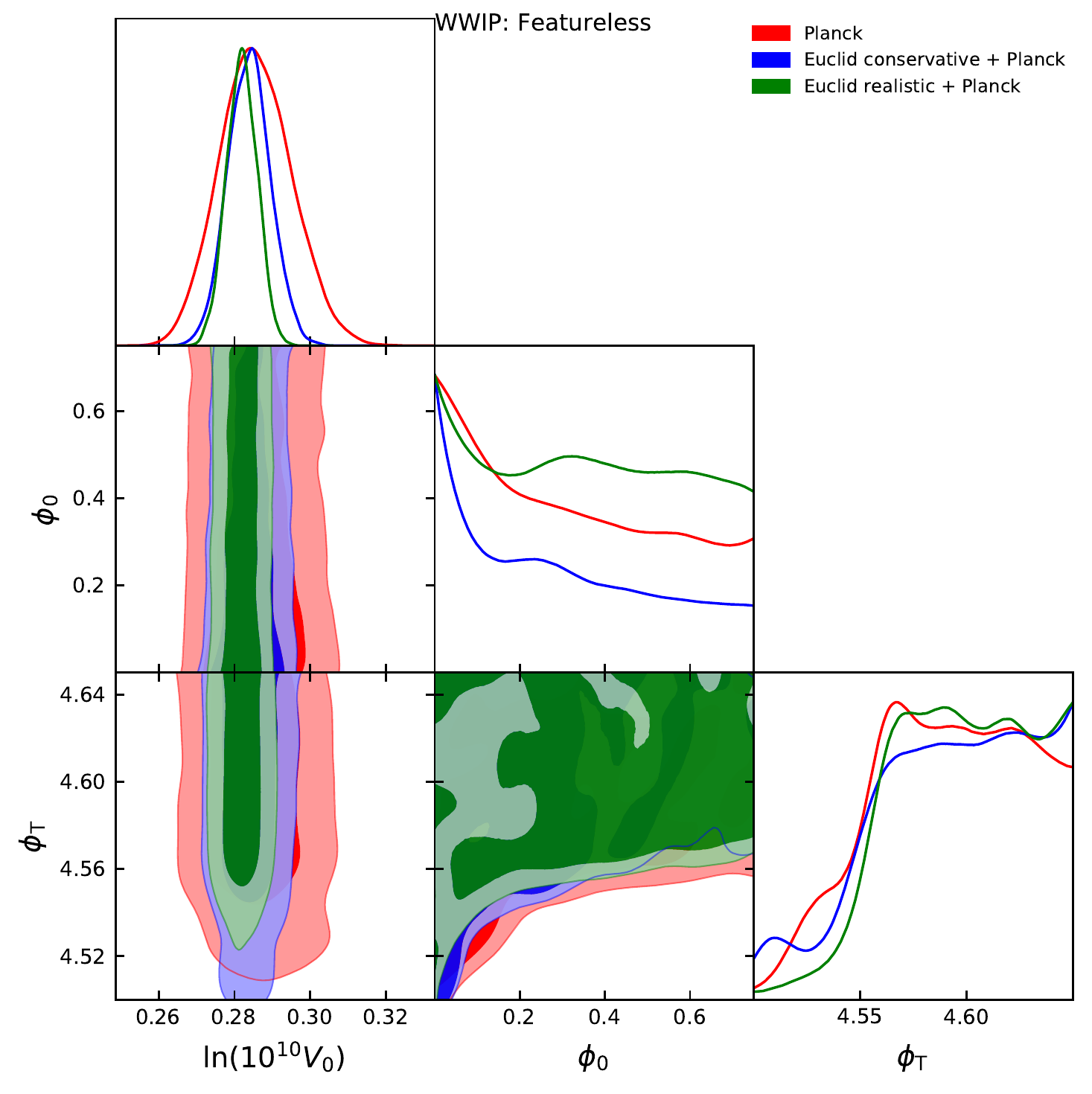}
	    \caption{One-dimensional posteriors and marginalized contours for the inflation parameters in the WWIP: Featureless model. Euclid Realistic shows some improvement over Euclid Conservative.
    \label{fig:WWIP1}}
\end{figure}

\begin{figure}[!htb]
\centering
	\includegraphics[width=\columnwidth]{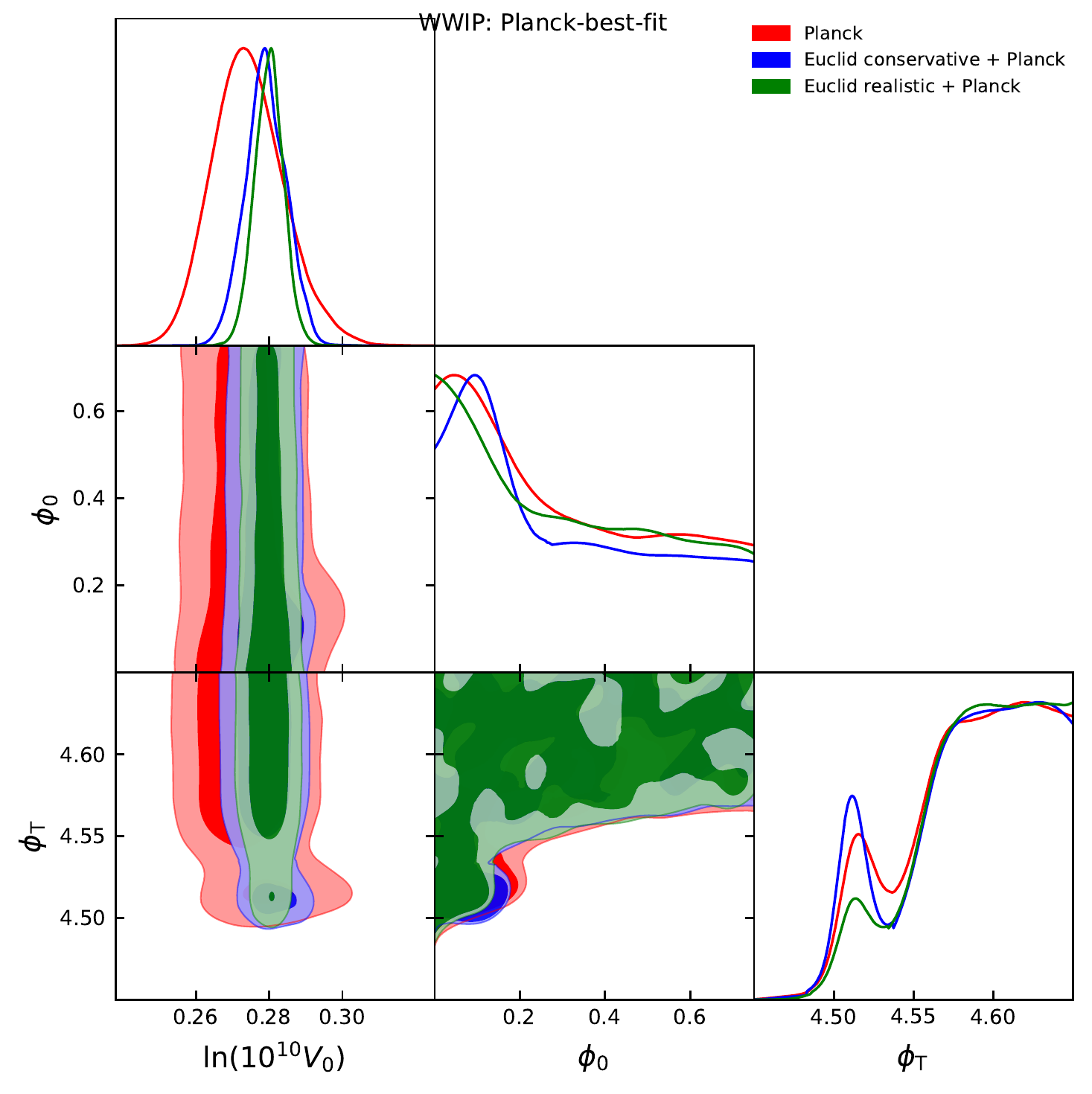}
    \caption{One-dimensional posteriors and marginalized contours for the inflation parameters in the WWIP: Planck-best-fit model. Again, Euclid Realistic shows some improvement over Euclid Conservative.
    \label{fig:WWIP2}}
\end{figure}

\begin{figure}[!htb]
\centering
	\includegraphics[width=\columnwidth]{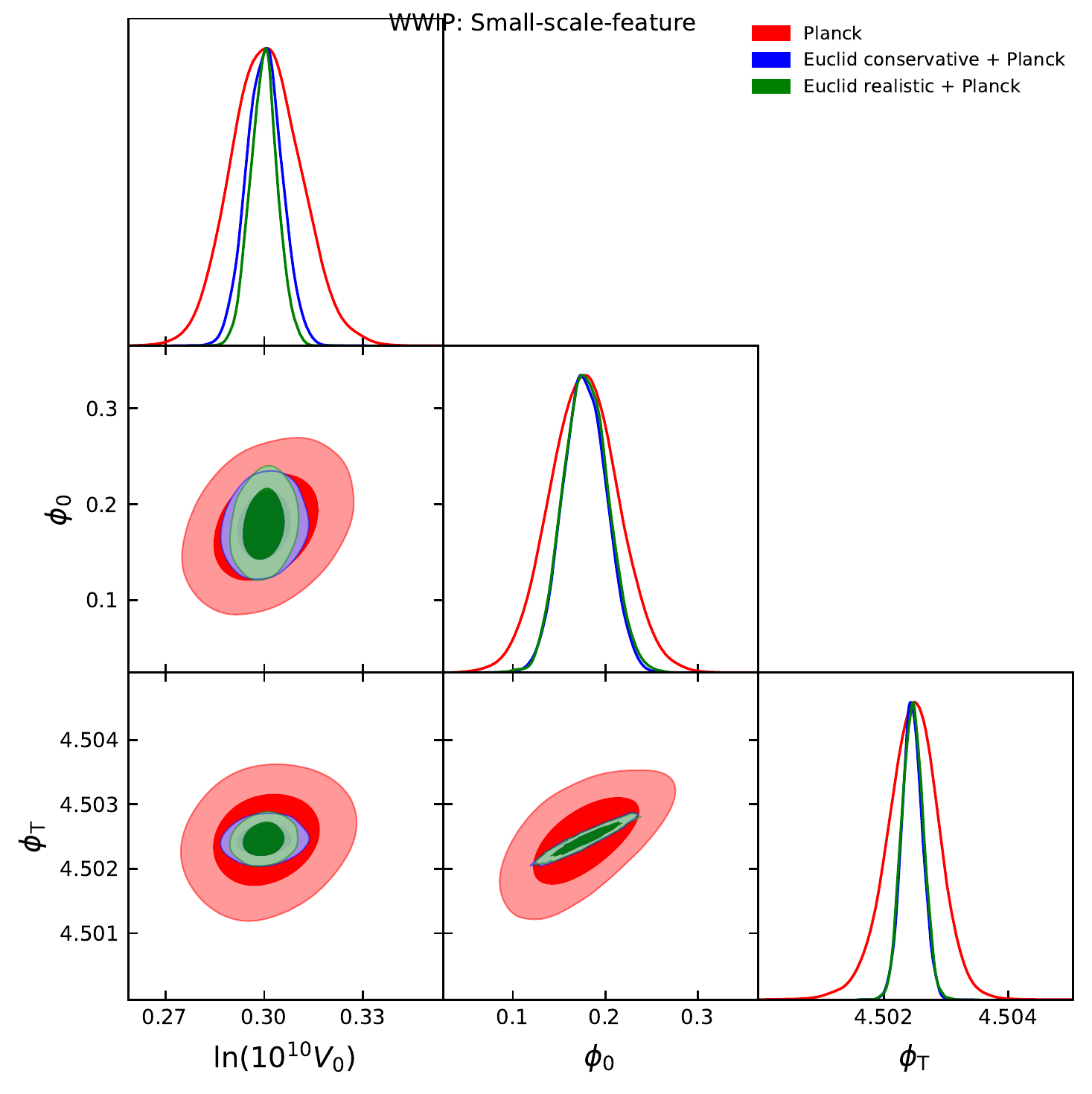}
    \caption{One-dimensional posteriors and marginalized contours for the inflation parameters in the WWIP: Small-scale-feature model. We obtain closed contours for all the inflation parameters, with a significant improvement with Euclid data are added. Euclid Realistic provides further improvement.
    \label{fig:WWIP3}}
\end{figure}

The WWIP potential has three parameters describing the primordial physics. The amplitude is set by $V_0$, while $\phi_0$ and $\phi_\mathrm{T}$ determine the transition in the potential and therefore the features. This model suppression at large scales, and wiggles throughout the primordial power spectrum, which gradually die out at small scales. 

Similarly to the WWI potential, we use a featureless fiducial generated with $\phi_0=0$ (called WWI: Featureless). We also use a model with the best fit to Planck temperature and polarization data \cite{Hazra2016}. We call it WWI: Planck-best-fit.
The third model is WWIP: Small-scale-feature, which has features extending towards even smaller scales ($k\sim0.2$ in $h\mathrm{Mpc}^{-1}$) in the primordial spectrum. The values of the inflation parameters in this model are not the best fit to Planck, but they are within 95 percent confidence limits.
%%%

%
The one-dimensional posteriors and marginalized contours for the inflationary potential parameters are shown in Figure~\ref{fig:WWIP1}--\ref{fig:WWIP3}. For all three models, we obtain $40$ to $50$ percent improvement in the constraints on $V_0$ when Euclid is added. 

For WWIP: Planck-best-fit, we find only marginal improvement with Euclid in $\phi_0$ (Figure~\ref{fig:WWIP2}). The current Planck best fit for WWIP has oscillations in the large to intermediate scales ($k\sim10^{-3} - 10^{-2}\mathrm{Mpc}^{-1}$). This range of scales is already well-probed by Planck. With Euclid we only expect to see marginal improvement. 

If future data support WWIP: Small-scale-feature, we can expect $40$ to $50$ per cent improvement in the constraints on $\phi_0$, leading to a detection of features with Euclid+Planck. This improvement is expected, since WWIP: Small-scale-feature has oscillations with higher magnitude than WWIP: Planck-best-fit, extending to smaller scales. These scales are accessible to Euclid. However, using the Realistic setup has no significant effect on the constraints for $\phi_0$ and $\phi_\mathrm{T}$, because the oscillations die out within the linear regime of the matter power spectrum.

The results for WWIP show that Euclid data can help improve constraints on inflationary parameters, but the improvement depends on the model parameter values. Since inflation features appear at particular scales, the scale probed by Euclid determines the improvements in constraints with respect to Planck CMB data. Wherever the wiggles are located at intermediate to small scales ($k\sim10^{-3} - 10^{-1}\mathrm{Mpc}^{-1}$), Euclid can play a significant role in detection when combined with Planck. Overall, the use of Euclid Realistic provides a sight improvement over Euclid Conservative. The most significant improvement is observed for $V_0$ (which determines the amplitude of the primordial power spectrum).
As we explain in ref.~\cite{Debono2020}, this is a consistent feature of the results for all models.
%%%

%
\section{Conclusions}
\label{sec:Conclusions}

The results in this paper extend the scope of ref.~\cite{Debono2020}, and are in qualitative agreement with other studies using other probes and other cosmological models \cite{Audren2013,Sprenger2018}. We already showed that the addition of Euclid data tightens the cosmological parameter constraints obtained by Planck alone. Here we show that the Realistic setup improves on the constraints from the Conservative setup. 

% Done
{The significance of the results for the Euclid mission were discussed in ref.~\cite{Debono2020}. In the present work, we can compare Euclid Conservative and Euclid Realistic, and assess the contribution of information from small scales, in the nonlinear regime of the matter power~spectrum.}

Euclid Realistic provides some improvement in constraints over Euclid Conservative for all parameters, both in the background cosmology and the inflation sector. This agrees with the general results obtained in ref.~\cite{Sprenger2018}.
Most of the information in the inflation sector comes from the cosmic microwave background, so the improvement of Euclid Realistic over Euclid Conservative is minimal for most of our WWI and WWIP models.
We observe the greatest improvement with WWI-D, which has high-frequency features going down to very small scales (see the power spectrum figures in ref.~\cite{Debono2020}). The Conservative $k$-cutoff discards most of the information from these very small scales. This suggests that the Euclid Realistic setup may enable us to probe inflation features of this kind. Improved modelling of the nonlinear regime would allow us to exploit information at small scales.

At large scales, we are limited by cosmic variance. At small scales, we are limited by theoretical errors. The open questions in cosmology are likely to be settled only by using data from multiple probes, and by exploiting their complementarity.

%%%

%%%

%%%%%%%%%%%%%%%%%%%%%%%%%%%%%%%%%%%%%%%%%%
\vspace{6pt}

%%%%%%%%%%%%%%%%%%%%%%%%%%%%%%%%%%%%%%%%%%
%\funding{}

%%%%%%%%%%%%%%%%%%%%%%%%%%%%%%%%%%%%%%%%%%
\acknowledgments

Ivan Debono acknowledges that the research work disclosed in this publication was funded up to 2019 by the 
REACH HIGH Scholars Programme -- Post-Doctoral Grants. The grant is part-financed by the European Union, Operational Programme II -- Cohesion Policy 2014--2020. 
The author gratefully acknowledges support from the CNRS/IN2P3 Computing Centre (Lyon -- France) for providing computing resources needed for this work.\\
This paper is a companion to another article published in 2020, co-authored by Ivan Debono, Dhiraj Kumar Hazra, Arman Shafieloo, George F. Smoot, and Alexei A. Starobinsky. The 2020 article contained the results for Planck and Euclid Conservative. The authors contributed to the software and the work of which the present paper is an extension.

%%%%%%%%%%%%%%%%%%%%%%%%%%%%%%%%%%%%%%%%%%

%%%%%%%%%%%%%%%%%%%%%%%%%%%%%%%%%%%%%%%%%%
%% optional

\vfill
\bibliography{Debono_MDPI_Proceedings}

\providecommand{\href}[2]{#2}\begingroup\raggedright\begin{thebibliography}{10}

\bibitem{WMAP:2003}
D.N.~{Spergel}, L.~{Verde}, H.V.~{Peiris}, E.~{Komatsu}, M.R.~{Nolta},
  C.L.~{Bennett} et~al., \emph{First-year {Wilkinson Microwave Anisotropy
  Probe} {(WMAP)} observations: {Determination} of cosmological parameters},
  \href{https://doi.org/10.1086/377226}{\emph{Astrophys. J, Supp.} {\bfseries
  148} (2003) 175}.

\bibitem{WMAP9}
G.~{Hinshaw}, D.~{Larson}, E.~{Komatsu}, D.N.~{Spergel}, C.L.~{Bennett},
  J.~{Dunkley} et~al., \emph{{Nine-year Wilkinson Microwave Anisotropy Probe
  (WMAP) Observations: Cosmological parameter results}},
  \href{https://doi.org/10.1088/0067-0049/208/2/19}{\emph{Astrophys. J, Supp.}
  {\bfseries 208} (2013) 19}.

\bibitem{Planck2018:params}
{Planck Collaboration VI}, \emph{{Planck 2018 results - {VI}. {Cosmological
  parameters}}},
  \href{https://doi.org/10.1051/0004-6361/201833910}{\emph{Astron. \&
  Astrophys.} {\bfseries 641} (2020) A6}.

\bibitem{Planck2018:inflation}
{Planck Collaboration X}, \emph{{Planck 2018 results - {X}. Constraints on
  inflation}}, \href{https://doi.org/10.1051/0004-6361/201833887}{\emph{Astron.
  \& Astrophys.} {\bfseries 641} (2020) A10}.

\bibitem{Debono2020}
I.~Debono, D.K.~Hazra, A.~Shafieloo, G.F.~Smoot and A.A.~Starobinsky,
  \emph{Constraints on features in the inflationary potential from future
  euclid data}, \href{https://doi.org/10.1093/mnras/staa1765}{\emph{Mon. Not.
  R. Astron. Soc.} {\bfseries 496} (2020) 3448}.

\bibitem{Hazra2014}
D.K.~Hazra, A.~Shafieloo, G.F.~Smoot and A.A.~Starobinsky, \emph{Wiggly whipped
  inflation}, \href{https://doi.org/10.1088/1475-7516/2014/08/048}{\emph{J.
  Cosmol. \& Astropart. Phys.} {\bfseries 8} (2014) 48}.

\bibitem{Kosowsky:1995}
A.~{Kosowsky} and M.S.~{Turner}, \emph{{CBR anisotropy and the running of the
  scalar spectral index}}, {\emph{Phys. Rev. D} {\bfseries 52} (1995) 1739}.

\bibitem{Bridle:2003}
S.L.~{Bridle}, A.M.~{Lewis}, J.~{Weller} and G.~{Efstathiou},
  \emph{{Reconstructing the primordial power spectrum}},
  \href{https://doi.org/10.1046/j.1365-8711.2003.06807.x}{\emph{Mon. Not. R.
  Astron. Soc.} {\bfseries 342} (2003) L72}.

\bibitem{Planck-Collaboration:2013aa}
{Planck Collaboration XVI}, \emph{{Planck 2013 results - XVI. Cosmological
  parameters}},
  \href{https://doi.org/10.1051/0004-6361/201321591}{\emph{Astron. \&
  Astrophys.} {\bfseries 571} (2014) A16}.

\bibitem{Planck-Collaboration2015aa}
{Planck Collaboration XIII}, \emph{{Planck 2015 results - {XIII}. Cosmological
  parameters}},
  \href{https://doi.org/10.1051/0004-6361/201525830}{\emph{Astron. \&
  Astrophys.} {\bfseries 594} (2016) A13}.

\bibitem{Hannestad_2001}
S.~Hannestad, \emph{Reconstructing the inflationary power spectrum from cosmic
  microwave background radiation data},
  \href{https://doi.org/10.1103/physrevd.63.043009}{\emph{Phys. Rev. D}
  {\bfseries 63} (2001) }.

\bibitem{Tegmark_2002}
M.~Tegmark and M.~Zaldarriaga, \emph{Separating the early universe from the
  late universe:{Cosmological parameter estimation beyond the black box}},
  \href{https://doi.org/10.1103/physrevd.66.103508}{\emph{Phys. Rev. D}
  {\bfseries 66} (2002) }.

\bibitem{Mukherjee_2003}
P.~Mukherjee and Y.~Wang, \emph{Model‐independent reconstruction of the
  primordial power spectrum from {Wilkinson Microwave Anistropy Probe} data},
  \href{https://doi.org/10.1086/379161}{\emph{Astrophys. J} {\bfseries 599}
  (2003) 1}.

\bibitem{Shafieloo_2004}
A.~Shafieloo and T.~Souradeep, \emph{Primordial power spectrum from {WMAP}},
  \href{https://doi.org/10.1103/physrevd.70.043523}{\emph{Phys. Rev. D}
  {\bfseries 70} (2004) }.

\bibitem{Kogo_2005}
N.~Kogo, M.~Sasaki and J.~Yokoyama, \emph{Constraining cosmological parameters
  by the cosmic inversion method},
  \href{https://doi.org/10.1143/ptp.114.555}{\emph{Progress of Theoretical
  Physics} {\bfseries 114} (2005) 555}.

\bibitem{Leach:2005av}
S.M.~Leach, \emph{Measuring the primordial power spectrum: {Principal component
  analysis of the cosmic microwave background}},
  \href{https://doi.org/10.1111/j.1365-2966.2006.10842.x}{\emph{Mon. Not. R.
  Astron. Soc.} {\bfseries 372} (2006) 646}.

\bibitem{TocchiniValentini:2005ja}
D.~Tocchini-Valentini, Y.~Hoffman and J.~Silk, \emph{{Non-parametric
  reconstruction of the primordial power spectrum at horizon scales from {WMAP}
  data}}, \href{https://doi.org/10.1111/j.1365-2966.2006.10031.x}{\emph{Mon.
  Not. R. Astron. Soc.} {\bfseries 367} (2006) 1095}.

\bibitem{Shafieloo_2008}
A.~Shafieloo and T.~Souradeep, \emph{{Estimation of primordial spectrum with
  post-WMAP 3-year data}},
  \href{https://doi.org/10.1103/physrevd.78.023511}{\emph{Phys. Rev. D}
  {\bfseries 78} (2008) }.

\bibitem{Nicholson_2009}
G.~Nicholson and C.R.~Contaldi, \emph{Reconstruction of the primordial power
  spectrum using temperature and polarisation data from multiple experiments},
  \href{https://doi.org/10.1088/1475-7516/2009/07/011}{\emph{J. Cosmol. \&
  Astropart. Phys.} {\bfseries 2009} (2009) 011}.

\bibitem{Paykari_2010}
P.~Paykari and A.H.~Jaffe, \emph{Optimal binning of the primordial power
  spectrum}, \href{https://doi.org/10.1088/0004-637x/711/1/1}{\emph{Astrophys.
  J} {\bfseries 711} (2010) 1}.

\bibitem{Gauthier_2012}
C.~Gauthier and M.~Bucher, \emph{Reconstructing the primordial power spectrum
  from the cmb}, \href{https://doi.org/10.1088/1475-7516/2012/10/050}{\emph{J.
  Cosmol. \& Astropart. Phys.} {\bfseries 2012} (2012) 050}.

\bibitem{Hlozek_2012}
R.~Hlozek, J.~Dunkley, G.~Addison, J.W.~Appel, J.R.~Bond, C.S.~Carvalho et~al.,
  \emph{The {Atacama Cosmology Telescope}: {A measurement of the primordial
  power spectrum}},
  \href{https://doi.org/10.1088/0004-637x/749/1/90}{\emph{Astrophys. J}
  {\bfseries 749} (2012) 90}.

\bibitem{Hazra_2013}
D.K.~Hazra, A.~Shafieloo and T.~Souradeep, \emph{Primordial power spectrum: a
  complete analysis with the wmap nine-year data},
  \href{https://doi.org/10.1088/1475-7516/2013/07/031}{\emph{J. Cosmol. \&
  Astropart. Phys.} {\bfseries 2013} (2013) 031}.

\bibitem{Dorn_2014}
S.~Dorn, E.~Ramirez, K.E.~Kunze, S.~Hofmann and T.A.~En{\ss}lin, \emph{Generic
  inference of inflation models by non-gaussianity and primordial power
  spectrum reconstruction},
  \href{https://doi.org/10.1088/1475-7516/2014/06/048}{\emph{J. Cosmol. \&
  Astropart. Phys.} {\bfseries 2014} (2014) 048}.

\bibitem{Hazra_2014}
D.K.~Hazra, A.~Shafieloo, G.F.~Smoot and A.A.~Starobinsky, \emph{Ruling out the
  power-law form of the scalar primordial spectrum},
  \href{https://doi.org/10.1088/1475-7516/2014/06/061}{\emph{J. Cosmol. \&
  Astropart. Phys.} {\bfseries 2014} (2014) 061}.

\bibitem{Hazra2014bJCAP}
D.K.~Hazra, A.~Shafieloo and T.~Souradeep, \emph{Primordial power spectrum from
  planck}, \href{https://doi.org/10.1088/1475-7516/2014/11/011}{\emph{J.
  Cosmol. \& Astropart. Phys.} {\bfseries 2014} (2014) 011}.

\bibitem{Hunt_2014}
P.~Hunt and S.~Sarkar, \emph{Reconstruction of the primordial power spectrum of
  curvature perturbations using multiple data sets},
  \href{https://doi.org/10.1088/1475-7516/2014/01/025}{\emph{J. Cosmol. \&
  Astropart. Phys.} {\bfseries 2014} (2014) 025}.

\bibitem{Hazra2014b}
D.K.~Hazra, A.~Shafieloo, G.F.~Smoot and A.A.~Starobinsky, \emph{Inflation with
  whip-shaped suppressed scalar power spectra},
  \href{https://doi.org/10.1103/PhysRevLett.113.071301}{\emph{Phys. Rev. Lett.}
  {\bfseries 113} (2014) 071301}.

\bibitem{Hazra2016}
D.K.~Hazra, A.~Shafieloo, G.F.~Smoot and A.A.~Starobinsky, \emph{Primordial
  features and {Planck} polarization},
  \href{https://doi.org/10.1088/1475-7516/2016/09/009}{\emph{J. Cosmol. \&
  Astropart. Phys.} {\bfseries 9} (2016) 009}.

\bibitem{Sprenger2018}
T.~Sprenger, M.~Archidiacono, T.~Brinckmann, S.~Clesse and J.~Lesgourgues,
  \emph{Cosmology in the era of {Euclid} and the {Square Kilometre Array}},
  \href{https://doi.org/10.1088/1475-7516/2019/02/047}{\emph{J. Cosmol. \&
  Astropart. Phys.} {\bfseries 2019} (2019) 047}.

\bibitem{Bingo}
D.K.~Hazra, L.~Sriramkumar and J.~Martin, \emph{{BINGO}: a code for the
  efficient computation of the scalar bi-spectrum},
  \href{https://doi.org/10.1088/1475-7516/2013/05/026}{\emph{J. Cosmol. \&
  Astropart. Phys.} {\bfseries 2013} (2013) 026}.

\bibitem{MontePython3}
T.~Brinckmann and J.~Lesgourgues, \emph{{MontePython 3: Boosted MCMC sampler
  and other features}},
  \href{https://doi.org/10.1016/j.dark.2018.100260}{\emph{Physics of the Dark
  Universe} {\bfseries 24} (2019) 100260}.

\bibitem{CLASS}
D.~Blas, J.~Lesgourgues and T.~Tram, \emph{The cosmic linear anisotropy solving
  system (class). part ii: Approximation schemes},
  \href{https://doi.org/10.1088/1475-7516/2011/07/034}{\emph{J. Cosmol. \&
  Astropart. Phys.} {\bfseries 2011} (2011) 034}.

\bibitem{Takahashi2012}
R.~{Takahashi}, M.~{Sato}, T.~{Nishimichi}, A.~{Taruya} and M.~{Oguri},
  \emph{Revising the {Halofit} model for the nonlinear matter power spectrum},
  \href{https://doi.org/10.1088/0004-637X/761/2/152}{\emph{Astrophys. J}
  {\bfseries 761} (2012) 152}.

\bibitem{Bird2012}
S.~{Bird}, M.~{Viel} and M.G.~{Haehnelt}, \emph{{Massive neutrinos and the
  non-linear matter power spectrum}},
  \href{https://doi.org/10.1111/j.1365-2966.2011.20222.x}{\emph{Mon. Not. R.
  Astron. Soc.} {\bfseries 420} (2012) 2551}.

\bibitem{Audren2013}
B.~Audren, J.~Lesgourgues, S.~Bird, M.G.~Haehnelt and M.~Viel, \emph{Neutrino
  masses and cosmological parameters from a euclid-like survey: Markov chain
  monte carlo forecasts including theoretical errors},
  \href{https://doi.org/10.1088/1475-7516/2013/01/026}{\emph{J. Cosmol. \&
  Astropart. Phys.} {\bfseries 1301} (2013) 026}.

\end{thebibliography}\endgroup

%%%%%%%%%%%%%%%%%%%%%%%%%%%%%%%%%%%%%%%%%%
\end{document}